\documentclass[12pt]{amsart}

\usepackage{amssymb}
\usepackage{amsthm}
\usepackage{amsbsy}
\usepackage{subfigure}
\usepackage{amsmath}
\usepackage{graphicx}

\usepackage{amsaddr}
\usepackage{comment}
\usepackage[usenames,dvipsnames]{xcolor}
\usepackage[margin = 1.1in]{geometry}
\usepackage{enumerate}
\usepackage[toc,page]{appendix}
\usepackage{lineno}
\usepackage{color}
\usepackage{hyperref}
\usepackage[ruled,vlined,linesnumbered]{algorithm2e}

\SetKwInOut{Parameter}{Parameters}

\definecolor{mygreen}{rgb}{0.1,0.75,0.2}

\numberwithin{equation}{section}

\DeclareMathOperator{\st}{s.t.~}

\newcommand{\la}{\langle}
\newcommand{\ra}{\rangle}
\newcommand{\pt}{\partial}
\newcommand{\eps}{\varepsilon}

\providecommand{\bbs}[1]{\left(#1\right)}
\newcommand{\aaa}[1]{\begin{equation}
begin{aligned} #1 \end{aligned}
\end{equation}}

\newcommand{\nnn}{\nonumber}
\newcommand{\ud}{\,\mathrm{d}}
\newcommand{\8}{\infty}

\newcommand{\sH}{\mathcal{H}}
\newcommand{\sP}{\mathcal{P}}
\newcommand{\bT}{\mathbb{T}}

\newcommand{\bR}{\mathbb{R}}

\newcommand{\fw}{{\text{fw}}}
\newcommand{\bw}{{\text{bw}}}
\begin{document}

\title[Enriched evolution of global sea surface height]{Enriched evolution of global sea surface height via generalized Schrodinger bridge and Fokker-Planck solver}

\author{Guangzhen Jin}
\address{Rosen Center for Advanced Computing, Purdue University, IN}
\email{jin456@purdue.edu}

\begin{abstract}
Global warming has been discussed for decades and is one of most popular topics in different areas of research. The sea level rise in recent decades, which was mainly caused by global warming, has drawn great attentions and interests from scientists because it’s crucial to human life as well as the entire earth system. A generalized Schr\"odinger bridge problem with an underlying energy landscape is used to model this process. We introduce an iterative numerical method for the associated mixed control problem with a given initial distribution (sea level height at the year 1994) and a given ending distribution (sea level height at the year 2014). The convergence of the introduced iterative method for finding the optimal transformation path of SSH is validated numerically. The evolution of sea level height from August 1994 to August 2014 has been characterized during the model simulation and the sea level height evolutions in several significant areas induced by ocean mesoscale eddies are reveled. 

\end{abstract}

%\section*{Acknowledge}
 
\maketitle

\section{Introduction}
Climate change, which refers to long-term shifts in temperatures and weather patterns of the earth, has been discussed for decades and is still one of the most popular topics in different areas of research. This change may be natural caused by the variations in the solar cycle. However, human activities have significantly impacted climate change since 1800s, primarily due to burning of fossil fuels such as oil, gas and coal. Known as the greenhouse effect, the emissions of greenhouse gas continue to rise and cause the temperature rising to a great extent. According to the historic records, the Earth is now about 1.1 Celsius degree warmer than it was in the late 1800s. And actually, the last decade (2011-2020) witnessed the warmest decade on record so it is also known as global warming. As a result, glaciers and ice sheets began to melt as a high speed. The seawater will also undergo thermal expansion as it warms. These effects will all cause the rising water level. As observed by Church and White (2011) \cite{CW2011}, the global mean sea level has risen about 8-9 inches (21-24 centimeters) since 1880. And in 2021, the global mean sea level was 3.8 inches above the 1993 levels, making it the highest annual average in the satellite record (1993-present). 

Global warming is the major cause for sea level rising through two ways. First of all, glaciers and ice sheets are melting as the temperature rose so they add water into the ocean. Second, the volume of the sea water expanded as the sea water temperature rose. 

Investigations on sea level change is one of the most important scientific topics because it significantly influences human life. In the United States, with more than $40\%$ of the US population live in coastal counties, almost $30\%$ of the population lives there. Globally, 8 of the world's 10 largest cities are near a coast. So, in urban settings along coastlines around the world, rising seas threaten infrastructure necessary for local jobs and regional industries. Roads, bridges, subways, water supplies, oil and gas wells, power plants, sewage treatment plants, landfills -  the list is practically endless - are all at risk from sea level rise. Moreover, high background water levels means that deadly and destructive storm surges, especially those caused by hurricanes.

As the development of ocean observation technologies, there have been two major methods measuring the sea level: tide gauges and satellite altimeters. One is in-situ observations and the other one is remote sensing observations. With the rapid expanding of super computers, different numerical models have also been applied to analyze and predict the sea level changes. Among all parameters, Sea Surface Height (SSH) is the most frequently used parameter to describe the global sea surface change. It is the height of the sea surface above a reference ellipsoid. This is the direct product of satellite altimetry. Sea surface height values are provided along the satellites’ ground tracks or at regular grids interpolated from the values determined along the satellite tracks. Instantaneous sea surface height values contain long-term, annual, seasonal and short-term temporal variations of the ocean surface.

Past and future sea level rise at specific locations on land may be more or less than the global average due to local factors: ground settling, upstream flood control, erosion, regional ocean currents, and whether the land is still rebounding from the compressive weight of Ice Age glaciers. In the United States, the fastest rates of sea level rise are occurring in the Gulf of Mexico from the mouth of the Mississippi westward, followed by the mid-Atlantic. Only in Alaska and a few places in the Pacific Northwest are sea levels falling, though that trend will reverse under high greenhouse gas emission pathways.

	In this paper, we aim to simulate an optimal transformation path from a given SSH at an initial time, say the year 1994, to the SSH at the year 2014. To find an optimal transformation path with only two given ending data.  A generalized Schr\"odinger bridge problem \cite{peyre2019computational}, which describes how to transform one distribution to another distribution with an underlying driving potential is used to model the transformation processes; see details in the next section. The Fokker-Planck solver based on the mixed optimal control algorithm is introduced to estimate the optimal path from initial density to final density. This algorithm is numerically shown to be stable and will be used to construct an optimal transformation path. The convergence of the introduced iterative method for finding the optimal transformation path of SSH is validated numerically. The process is then applied to the analysis of the evolution process of the global SSH within the past 20 years (from 1994 to 2014). The evolution process in the North Pacific is analyzed in detail to illustrate the rapid change of SSH under the background of global warming.

\section{Background in generalized Schr\"odinger bridge problem}

\subsection{The Schrodinger bridge problem and optimal control formulation}
The Schr\"odinger bridge problem (SB) was first introduced by  Schr\"odinger in 1932 \cite{Schrodinger1932theorie} and now becomes a fundamental framework in the physics, mathematics, engineering and information geometry to modeling the ensemble statistical transition path between fixed initial and final distributions.

Given initial density $\rho_0$ and final density $\rho_1$, denote (SB) as the Schr\"odinger bridge problem
\begin{align}
&\min_{b,\rho} \frac12 \int \int |b|^2 \rho \ud x \ud t\\
& \st \pt_t \rho + \nabla \cdot(\rho b)= \gamma \Delta \rho, \quad \rho_0 = \mu, \,\, \rho_1 = \nu. \nnn
\end{align}
When $\gamma\to 0$, the optimal $b$ is the optimal velocity field in Optimal transport.

\subsection{Generalized Schr\"odinger Bridge problem with an energy landscape}
 
It worth to notice the above Schr\"odinger Bridge problem can be generalized to include an     energy functional $E$ beyond the entropy. 
An internal energy showing an underlying driving potential can be used to guide the evolution of SSH patterns, which may be affected by multiple factors as a result of global warming. The combined driven force of those factors is usually vague and difficult to distinguish from each other. Thus the art of choosing proper underlying driving potential is case by case.

For simplicity, we consider distributions in the probability space $\sP$ on $\bT^d$ or $\bR^d$.
 Denote the final distribution of SSH as $\pi$. Based on the Boltzmann analysis, we can regard the final distribution as a probability that is proportional to the exponential of a potential difference
 \begin{equation}
 \pi \propto e^{-\frac{1}{\gamma} U}.
 \end{equation}
   Then we use the     relative entropy as a typical internal energy
\begin{equation}
E = \gamma \int \rho \ln \frac{\rho}{\pi} \ud x, \quad \pi \propto e^{-\frac{1}{\gamma} U}.
\end{equation}
It is easy to calculate the first variation of $E$ is
\begin{equation}
\frac{\delta E}{\delta \rho} = \gamma \ln\frac{\rho}{\pi} + \gamma = \gamma \ln \rho + U + \gamma.
\end{equation}
One have
the general Schr\"odinger bridge problem (SBg)
\begin{align}\label{SBg}
&\min_{b,\rho} \frac12 \int_0^1 \int |b|^2 \rho \ud x \ud t\\
& \st \pt_t \rho + \nabla \cdot(\rho b)=   \nabla \cdot \bbs{ \rho \nabla \frac{\delta E}{\delta \rho}}, \quad \rho_0 = \mu, \,\, \rho_1 = \nu. \nnn
\end{align}
It is equivalent to
the so called general Yasue problem \cite{yasue1981stochastic} 
\begin{align}
& \min  \int_0^1\int  \frac12|v|^2 \rho \ud x \ud t + \frac{1}{2} |\nabla  \frac{\delta E}{\delta \rho} |^2 \rho \ud x \ud t\\
& \st  \pt_t \rho + \nabla \cdot (\rho v)=0, \quad  \rho_0 = \mu, \,\, \rho_1 = \nu. \nnn
\end{align}
Indeed, this can be seen by the changing variable $b=v + \nabla \frac{\delta E}{\delta \rho}$ in \eqref{SBg}. Then one can directly compute the running cost with Lagrangian $L$, 
\begin{align*}
L:=&\frac12\int |b|^2 \rho \ud x \\
=& \int \bbs{\frac12 |v|^2 \rho + \frac12 |\nabla \frac{\delta E}{\delta \rho}|^2 \rho + \rho v \cdot \nabla  \frac{\delta E}{\delta \rho} } \ud x\\
=& \int \bbs{\frac12 |v|^2 \rho + \frac12 |\nabla \frac{\delta E}{\delta \rho}|^2 \rho + \pt_t \rho \frac{\delta E}{\delta \rho} } \ud x,
\end{align*}
where we used the constraint $\pt_t \rho + \nabla \cdot (\rho v)=0$.
Notice the last term above is $\frac{\ud E}{\ud t}$ which after the time integration is a constant depending only on the fixed initial distribution and the ending distribution.

We remark that there is another way to recast the corresponding Lagrangian function $L$  as
\begin{align}
L:=&\frac12\int |b|^2 \rho \ud x = \frac14 \|\nabla \cdot (\rho b)\|_{H^{-1}(\rho)}^2\\
=& \frac14 \|\pt_t \rho - \nabla \cdot \bbs{\rho \nabla \frac{\delta E}{\delta \rho}}\|^2_{H^{-1}(\rho)}.
\end{align}
This $L$ indeed measures the residual of gradient flow of $E$ in the $H^{-1}(\rho)$ norm. This $L$ is also the $L$-function in the Large derivation principle of the average process $X_\eps = \frac{1}{N}X_i = \gamma X_i$, with $X_i$ being the i.i.d Markov process with generator $Qf=\Delta f - \nabla U \cdot \nabla f$ \cite{FW, gao2022transition, gao2022revisit}.

In the next section, we will use the generalized Schr\"odinger Bridge problem with various energy landscape to design algorithms in image morphing and transition state theory. 

\subsection{Hopf-Cole transformation}
In this section, we convert the optimization problem \eqref{SBg} as a coupled PDE system with two boundary conditions.

First we derive the Euler-Langrange-Bellman equation for the generalized Schr\"odinger bridge problem \eqref{SBg}.

One can derive the Euler-Lagrange equations using Lagrangian multiplier $\phi_t(x)$
 \begin{equation}
 \sup_{b_t,\rho_t} \inf_{\Phi_s}  \bbs{  \int_0^1 \bbs{ \int  \bbs{\frac12|b|^2 \rho_t(x) +   \phi_t(x)[ \pt_t \rho_t + \nabla\cdot(\rho_t  b_t) -   \nabla \cdot \bbs{\rho \nabla \frac{\delta E}{\delta \rho}}]}   \ud x      } \ud s }.  
 \end{equation}
 Notice $\nabla \frac{\delta E}{\delta \rho} =   \gamma\frac{\nabla \rho}{\rho} + \nabla U$. Then the constraint becomes the Fokker-Planck equation for the drift-diffusion process
 \begin{equation}
 \pt_t \rho_t + \nabla\cdot(\rho_t  b_t) =   \nabla \cdot \bbs{\gamma \nabla \rho + \rho \nabla U }.
 \end{equation}
 Thus the Euler-Lagrange equations are
 \begin{equation}\label{mean_EL}
 \begin{aligned}
   \pt_t \rho_t + \nabla\cdot(\rho_t  b_t) =   \nabla \cdot \bbs{\gamma \nabla \rho + \rho \nabla U },\\
  \frac12|b|^2 - \pt_t \phi_t -b \cdot \nabla \phi +   \nabla \phi \cdot \nabla U - \gamma  \Delta \phi   =0,\\
  b(x) -   \nabla \phi_t=0.\\
 \end{aligned}
 \end{equation}

Therefore,  the optimal control velocity is given by 
$$b_t(x)= \nabla \phi_t(x)$$ 
and the Euler-Lagrange-Bellman equation becomes
\begin{equation}\label{ELB}
\begin{aligned}
\pt_t \phi + \frac12 |\nabla \phi|^2  -   \nabla \phi \cdot \nabla U + \gamma  \Delta \phi = 0,
\\
\pt_t \rho_t + \nabla \cdot \bbs{\rho_t (\nabla \phi_t-  \nabla U)}=\gamma  \Delta \rho,\\
\rho_{t=0} = \mu, \quad \rho_{t=1}  =\nu.
\end{aligned}
\end{equation}

Now we introduce a Hamiltonian 
functional $\sH:   \sP(\bT^d) \times C^1(\bT^d)\to \bR$
\begin{equation}\label{functionalHH}
\sH(\rho(\cdot), \phi(\cdot)):=  \int_{\bT^d}  \bbs{\frac12 |\nabla \phi|^2 - \la \nabla \phi, \nabla \frac{\delta E}{\delta \rho} \ra} \rho(x) \ud x.
\end{equation}
Then by elementary computations, we have
\begin{align*}
\frac{\delta \sH}{\delta \phi}(\rho, \phi) = -\nabla \cdot \bbs{\rho \nabla \phi} + \nabla \cdot \bbs{\rho \nabla \frac{\delta E}{\delta \rho}},\\
\frac{\delta \sH}{\delta \rho}(\rho, \phi) = \frac12|\nabla \phi|^2 + \gamma \Delta \phi - \nabla \phi \cdot \nabla U.
\end{align*} 
Thus the Euler-Lagrange-Bellman equation \eqref{ELB} becomes a Hamiltonian system in infinite dimensional space
\begin{equation}
\begin{aligned}
\pt_t \rho = \frac{\delta \sH}{\delta \phi}(\rho, \phi),\\
\pt_t \phi  = - \frac{\delta \sH}{\delta \rho}(\rho, \phi).
\end{aligned}
\end{equation}

Then by the Hopf-Cole transformation (c.f. \cite{Ishii_Evans_1985, gao2021large, gao2022transition}): $(\eta^\bw, \eta^\fw) \to (\rho , \phi)$ defined as
\begin{align}\label{HC}
\delta E (\rho) = \delta E(\eta^\bw) + \delta E(\eta^\fw), \\
\phi = 2 \delta E(\eta^{\bw}),
\end{align}
where $\delta E (\rho) = \gamma \log \rho + \gamma + U $ is the first variation of $E$.

%\begin{align*}
%\delta E (\rho) = \delta E(\eta^\bw) + \delta E(\eta^\fw), \\
%S = \phi - \delta E(\rho) =   \delta E(\eta^{\bw})-\delta E(\eta^\fw).
%\end{align*}

Then we have 
\begin{align}
\eta^\bw = e^{\frac{\phi}{2\gamma}} e^{-\frac{U}{\gamma} - 1}, \quad \eta^\fw = \rho e^{-\frac{\phi}{2\gamma}}.
\end{align}
Or equivalently
\begin{align}
\phi = 2 \gamma \ln \eta^\bw + 2 U + 2\gamma,\\
\rho = e^{\frac{\phi}{2\gamma}} \eta^\fw = \eta^\bw \eta^\fw e^{\frac{U}{\gamma}+1}.
\end{align}

With this transformation, one can check
\begin{align*}
\frac{\nabla \phi}{2} = \gamma \frac{\nabla \eta^\bw}{\eta^\bw} + \nabla U,\\
\gamma \frac{\nabla \rho}{\rho} - \frac{\nabla \phi}{2} = \gamma \frac{\nabla \eta^\fw}{\eta^\fw}.
\end{align*}
Then in terms of the $(\eta^\fw, \eta^\bw)$ the ELB equations \eqref{ELB} are
\begin{equation}
\begin{aligned}
\pt_t \eta^\bw + \gamma \Delta \eta^\bw + \nabla \eta^\bw \cdot \nabla U +  \eta^\bw  \Delta U =0,\\
-\pt_t \eta^\fw + \gamma \Delta \eta^\fw + \nabla \eta^\fw \cdot \nabla U +  \eta^\fw  \Delta U =0.
\end{aligned}
\end{equation}
or equivalently
\begin{equation}
\begin{aligned}
\pt_t \eta^\bw + \gamma \Delta \eta^\bw + \nabla \cdot(\eta^\bw  \nabla U) =0,\\
-\pt_t \eta^\fw + \gamma \Delta \eta^\fw + \nabla \cdot(\eta^\fw  \nabla U) =0.
\end{aligned}
\end{equation}
Define the Fokker-Planck operator
\begin{equation}
FP^U(\eta) := \gamma \Delta \eta + \nabla \cdot (\eta \nabla U) = \nabla \cdot \bbs{\eta (\gamma \nabla \ln \eta + \nabla U)}.
\end{equation}
With a given equilibrium $\pi=e^{-U}$, denote $m:=\frac{1}{\gamma}$ and $\pi^m=e^{-\frac{U}{\gamma}}$.
With $\pi^m= e^{-\frac{U}{\gamma}}$, one can rewrite the FP operator as
\begin{equation}
\begin{aligned}
FP^{\pi^m}(\eta) :=& \gamma \Delta \eta + \nabla \cdot (\eta \nabla U)= \gamma \nabla \cdot \bbs{\eta \bbs{ \nabla \ln \eta + \frac{1}{\gamma} \nabla U}} \\
=& \gamma \nabla \cdot \bbs{\eta \nabla \ln \frac{\eta}{\pi^m}}= \gamma \nabla \cdot \bbs{\pi^m \nabla \frac{\rho}{\pi^m}}.
\end{aligned}
\end{equation}

Therefore the $(\eta^\fw, \eta^\bw)$ equations
become 
\begin{equation}
\begin{aligned}
-\pt_t \eta^\bw = FP^U(\eta^\bw), \quad \eta^\bw|_{t=1}=\eta^\bw_1, \\
\pt_t \eta^\fw = FP^U (\eta^\fw), \quad \eta^\fw|_{t=0}= \eta^\fw_0.
\end{aligned}
\end{equation}

The gSB problem is to find the data $\eta^\fw_0$ and $\eta^\bw_1$ such that 
\begin{align}
\rho_0=\eta^\fw_0 \eta^\bw(0)e^{\frac{U}{\gamma}+1},\\
\rho_1 =\eta^\fw(1)  \eta^\bw_1 e^{\frac{U}{\gamma}+1}.
\end{align}
As a result, the obtained $\eta^\fw(t)$ and $\eta^\bw(t)$ can be used to construct the gSB solution
\begin{equation}
\rho(t) = \eta^\fw(t) \eta^\bw(t) e^{\frac{U}{\gamma}+1}.
\end{equation}

The Fokker-Planck equation can be solved via the finite volume method introduced in \cite{scharfetter1969large, eymard2000finite, gao2023data, gao2021inbetweening}, which can also be generalized to the solver for Fokker-Planck equations on a manifold. The first difficulty here is the initial data and the ending data for $\eta^{\fw}$ and for $\eta^{bw}$ are unknown. Thus some iterative shooting method for finding $\eta^{\fw}$ and   $\eta^{bw}$ based on given $\rho_0$ and $\rho_1$ are necessary. Some existing ones are \cite{sinkhorn1964relationship, cuturi2013sinkhorn}. On the other hand, how to design the potential $U$ in the forward and the backward equations is an art based on the application in the current context. We refer to \cite{gao2018new, yg20, gao2022revisit} for various design of either a reversible or an irreversible drift to adjust the optimally controlled trajectories in the Fokker-Planck equation.

\subsection{Double control problem}
\begin{figure}
 \includegraphics[scale=0.5]{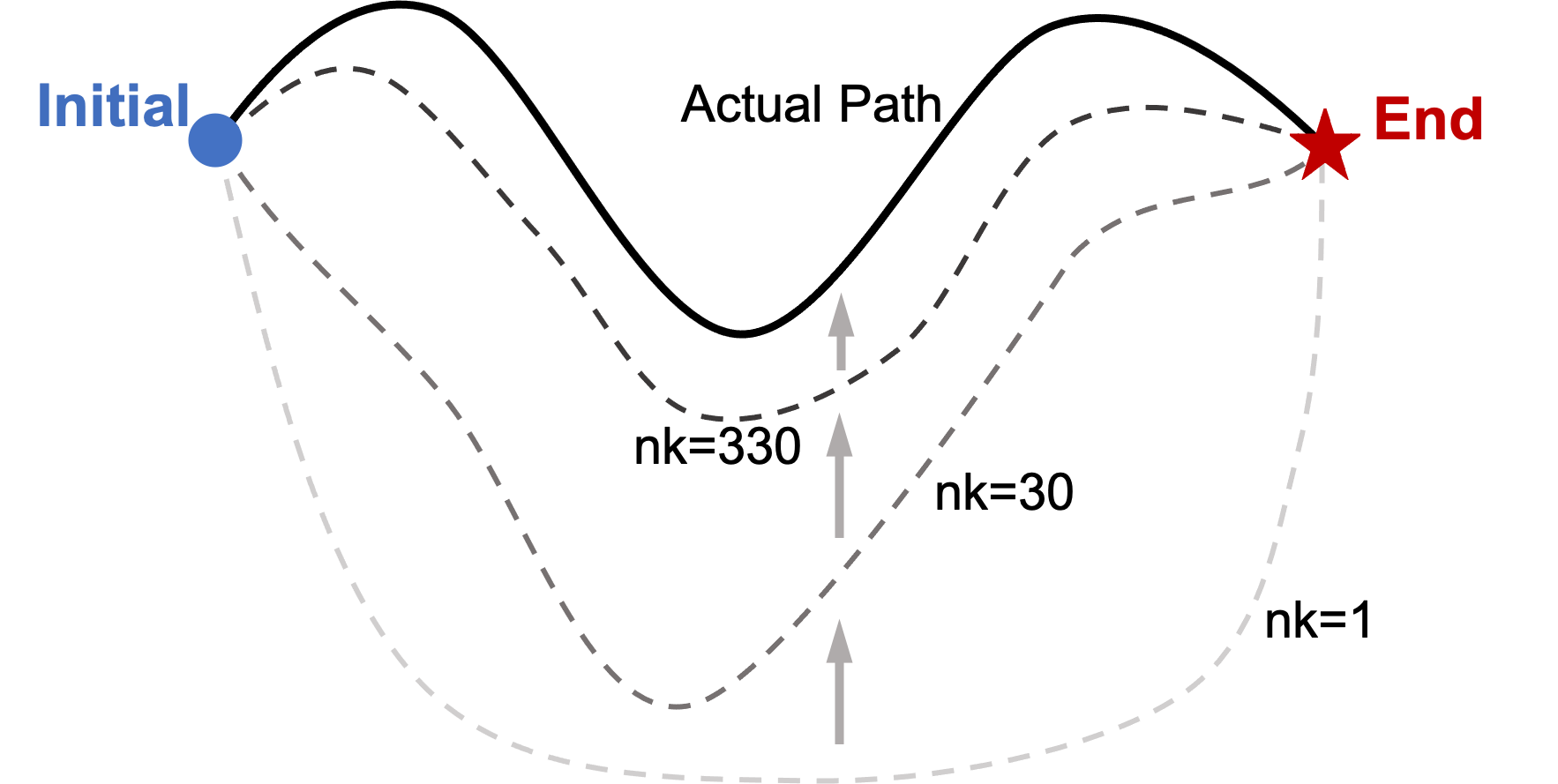} 
 \caption{The sketch showing the optimal path evolution as the iteration goes on.}\label{fig5}
 \end{figure}

In this section, we introduce the double control method for computing the optimal transformation path. The idea is illustrated in figure \ref{fig5}.
 
 Given the  initial/ending distributions $p_0 = \mu, \,\, p_1 = \nu$, one first solves the forward control problem with the driving potential $U_1= -\gamma \log \nu.$ Then with the  same initial/ending distributions $q_0 = \mu, \,\, q_1 = \nu$, one  solves the backward control problem with the driving potential $U_0= -\gamma \log \mu.$
 
Precisely, define the forward control problem
\begin{align}\label{SBf}
\min_{u,p}& \frac12 \int \int |u|^2 p \ud x \ud t\nnn\\
 \st &\pt_t p + \nabla \cdot(p u)=   \nabla \cdot \bbs{ p \nabla \frac{\delta E}{\delta p}} = \gamma \Delta p + \nabla \cdot \bbs{p \nabla U_1}\\
 & p_0 = \mu, \,\, p_1 = \nu. \nnn
\end{align}
Define the backward control problem
\begin{align}\label{SBb}
\min_{v,q}& \frac12 \int \int |v|^2 q \ud x \ud t\nnn\\
 \st &-\pt_t q + \nabla \cdot(q v)=   \nabla \cdot \bbs{ q \nabla \frac{\delta E}{\delta q}} = \gamma \Delta q + \nabla \cdot \bbs{q \nabla U_0}\\
 & q_0 = \mu, \,\, q_1 = \nu. \nnn
\end{align}

Then the Euler-Lagrange-Bellman equation to the forward control problem \eqref{SBf} is
\begin{equation}
\begin{aligned}
\pt_t \phi + \frac12 |\nabla \phi|^2 + \gamma \Delta \phi - \nabla \phi \cdot \nabla U_1 = 0,
\\
\pt_t  p + \nabla \cdot \bbs{p (\nabla \phi-\nabla U_1)}=\gamma \Delta  p
\end{aligned}
\end{equation}
with $u= \nabla \phi$.
And the Euler-Lagrange-Bellman equation to the backward control problem \eqref{SBb} is
\begin{equation}
\begin{aligned}
-\pt_t \varphi + \frac12 |\nabla \varphi|^2 + \gamma \Delta \varphi - \nabla \varphi \cdot \nabla U_0 = 0,
\\
-\pt_t  q + \nabla \cdot \bbs{q (\nabla \varphi-\nabla U_0)}=\gamma \Delta  q
\end{aligned}
\end{equation}
with $v= \nabla \varphi$.

Under the same Hopf-Cole transformation \eqref{HC} for $(p,\phi)$,
\begin{align}
\phi = 2 \gamma \ln \eta^\bw + 2 U_1 + 2\gamma,\\
p = e^{\frac{\phi}{2\gamma}} \eta^\fw= \eta^\bw \eta^\fw  e^{\frac{U_1}{\gamma}+1},
\end{align}
we have 
 the $(\eta^\fw, \eta^\bw)$ equations
\begin{equation}
\begin{aligned}
-\pt_t \eta^\bw = FP^{U_1} (\eta^\bw), \quad \eta^\bw(1)=\eta^\bw_1, \\
\pt_t \eta^\fw = FP^{U_1} (\eta^\fw), \quad \eta^\fw(0)=\eta^\fw_0.
\end{aligned}
\end{equation}
Then forward control problem \eqref{SBf} is to find the data $\eta^\fw_0$ and $\eta^\bw_1$ such that 
\begin{align}
p_0=\eta^\bw(0)\eta^\fw_0 e^{\frac{U_1}{\gamma}+1},\\
p_1 = \eta^\bw_1\eta^\fw(1) e^{\frac{U_1}{\gamma}+1}.
\end{align}
Particularly, when $p_1= e^{-\frac{U_1}{\gamma}-1}$, we have $1=\eta_1^\bw \eta^\fw(1).$ Thus the data for $\eta$ equation is
\begin{align}
p_0 p_1  =\eta^\bw(0)\eta^\fw_0,\\
p_1^2 = \eta^\bw_1\eta^\fw(1).
\end{align}
As a result, the obtained $\eta^\fw(t)$ and $\eta^\bw(t)$ can be used to construct the  solution to the forward control problem \eqref{SBf}
\begin{equation}
p(t)p_1 =\eta^\bw(t)\eta^\fw(t) .
\end{equation}

Similarly,
under the  Hopf-Cole transformation \eqref{HC} for $(q,\varphi)$,
\begin{align}
\varphi = 2 \gamma \ln \xi^\fw + 2 U_0 + 2\gamma,\\
q = e^{\frac{\varphi}{2\gamma}} \xi^\bw= \xi^\bw \xi^\fw  e^{\frac{U_0 }{\gamma}+1},
\end{align}
we have 
 the equations for $(\xi^\fw, \xi^\bw)$
\begin{equation}
\begin{aligned}
\pt_t \xi^\fw = FP^{U_0} (\xi^\fw), \quad \xi^\fw(0)=\xi^\fw_0, \\
-\pt_t \xi^\bw = FP^{U_0} (\xi^\bw), \quad \xi^\bw(1)=\xi^\bw_1.
\end{aligned}
\end{equation}
Then backward control problem \eqref{SBb} is to find the data $\xi^\fw_0$ and $\xi^\bw_1$ such that 
\begin{align}
q_0=\xi^\bw(0)\xi^\fw_0 e^{\frac{U_0}{\gamma}+1},\\
q_1 = \xi^\bw_1\xi^\fw(1) e^{\frac{U_0}{\gamma}+1}.
\end{align}
Particularly, when $q_0= e^{-\frac{U_0}{\gamma}-1}$, we have $q_0^2=\xi_0^\fw \xi^\bw(0).$
Thus the data for $\xi$ equation is
\begin{align}
q_0^2=  \xi^\bw(0) \xi^\fw_0 ,\\
q_1 q_0 =   \xi^\bw_1 \xi^\fw(1) .
\end{align}

As a result, the obtained $\xi^\fw(t)$ and $\xi^\bw(t)$ can be used to construct the  solution to the forward control problem \eqref{SBb}
\begin{equation}
q(t) q_0= \xi^\bw(t)\xi^\fw(t) .
\end{equation}

\section{Numerical schemes for a mixed control problem}
In this section, based on the description for the forward/backward control problem in the last section, we introduce the mixed control problem and give the associated algorithm for this mixed control problem. 

Take the driving potential as $U_1= -\gamma \log \nu$ in the forward Fokker-Planck equation and take  the driving potential $U_0= -\gamma \log \mu$ in the backward Fokker-Planck equation.  Then we find solutions with proper initial $\eta_0^\fw$ and ending data $\xi_1^\bw$ to
\begin{align}
\pt_t \eta^\fw = FP^{U_1} (\eta^\fw), \quad \eta^\fw(0)=\eta^\fw_0, \label{FP1}\\
-\pt_t \xi^\bw = FP^{U_0} (\xi^\bw), \quad \xi^\bw(1)=\xi^\bw_1. \label{FP2}
\end{align}
We also need to find $\eta^\fw_0$ and $\xi_1^\bw$ such that
\begin{align}
p_0^2 = \xi^\bw(0) \eta_0^\fw  , \label{i1}\\
 p_1^2= \xi_1^\bw \eta^\fw(1)  .\label{i2}
\end{align}

As a result, the obtained $\eta^\fw(t)$ and $\xi^\bw(t)$ can be used to construct 
\begin{equation}
\rho(t)^2 = \eta^\fw(t) \xi^\bw(t), \quad 0\leq t\leq 1.
\end{equation}

\subsection{Mixed control scheme via a generalized Sinkhorn algorithms}
We now design an algorithm based on the generalized Sinkhorn algorithms \cite{sinkhorn1964relationship}. The convergence of this algorithm is an interesting question which need some analysis. In the example for the evolution of SSH, we numerically show the convergence of this iteration method.

Denote the numerical solution at $k$-th iteration as $\eta^{\fw,k}, \eta^{\bw,k}$. Below we describe the algorithm.
~~\\
Input: Begin image $p_0$, End image $p_1$, Initial guess $\eta^{\fw,0}_0$.\\
Step 1. Given data $\eta^{\fw,k}_0$, solve FP \eqref{FP1} to time $t=1$. Denote the obtained solution as $\eta^{\fw,k}(1)$. \\
Step 2. Based on \eqref{i2}, solve $\xi^{\bw, k}_1$ by
\begin{equation}
p_1^2 =  \xi^{\bw, k}_1 \eta^{\fw,k}(1) .
\end{equation}
Step 3. Using data $\xi^{\bw, k}_1$, solve FP \eqref{FP2} to time $t=0$. Denote the obtained solution as $\xi^{\bw,k}(0).$
\\
Step 4. Based on \eqref{i1}, solve $\eta^{\fw, k+1}_0$ by
\begin{equation}
p_0^2 =  \xi^{\bw, k}(0) \eta^{\fw,k+1}_0.
\end{equation}
Output: The converged $\eta_0^{\fw}:=\lim_{k\to \8} \eta_0^{\fw,k}$, $\xi_1^{\bw}:=\lim_{k\to \8} \xi_1^{\bw,k}$. Then with these $\eta_0^\fw, \xi^{\bw}_1$ solve  $\eta^\fw(t)$,$\xi^\bw(t)$ and construct 
\begin{equation}
\rho(t)^2 = \eta^\fw(t) \xi^\bw(t) , \quad 0\leq t\leq 1.
\end{equation}

\section{Experiment for SSH evolution}

\subsection{Data}
In this paper, the global SSH data are derived from the HYCOM (HYbrid Coordinate Ocean Model) re-analysis dataset. The HYCOM consortium is a multi-institutional effort sponsored by the National Ocean Partnership Program (NOPP), as part of the U. S. Global Ocean Data Assimilation Experiment (GODAE), to develop and evaluate a data-assimilative hybrid isopycnal-sigma-pressure (generalized) coordinate ocean model. The GODAE objectives of three-dimensional depiction of the ocean state at fine resolution in real time, provision of boundary conditions for coastal and regional models, and provision of oceanic boundary conditions for a global coupled ocean-atmosphere prediction model, are being addressed by a partnership of institutions that represent a broad spectrum of the oceanographic community.

HYCOM is a product from the HYCOM Consortium that forecasts global ocean conditions.  The HYCOM $+$ NCODA Global $1/12$ degree Analysis (GLBy0.08/expt\_93.0) provides five variables including Sea Surface Height as well as eastward velocity, northward velocity, in-situ temperature, and salinity at 40 vertical depth levels ranging from 0 m at the ocean surface to 5,000 m. It combines the model results and satellite observation data with the horizontal grid to be 0.08 degree along longitude and 0.08 degree along latitude. 

\begin{figure}
\includegraphics[scale=0.18]{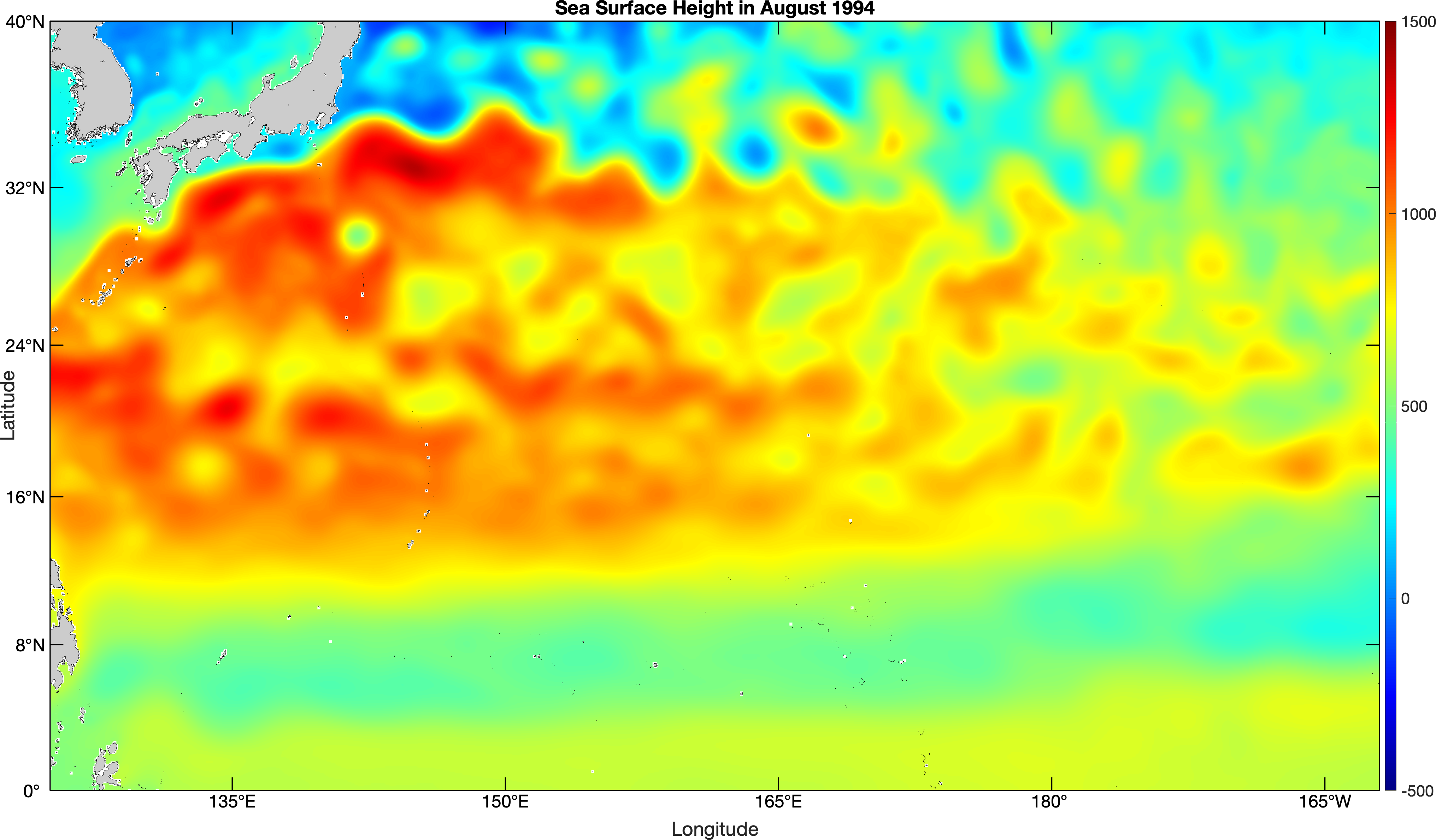} 
\caption{The monthly mean sea surface height in August, 1994. It is used as the initial distribution $\rho_0$.}\label{fig1}
\end{figure}

\begin{figure}
\includegraphics[scale=0.18]{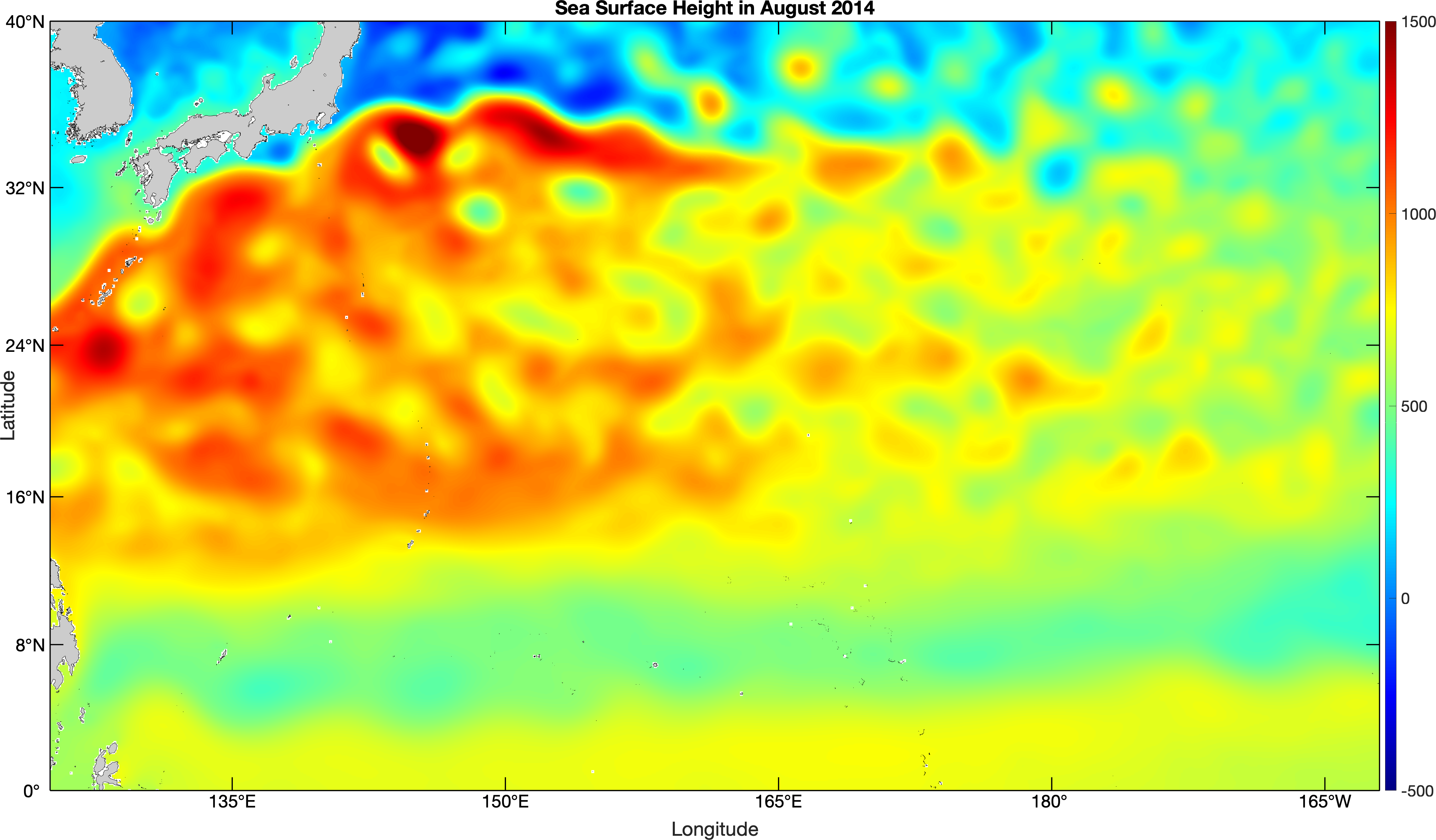} 
\caption{The monthly mean sea surface height in August, 2014. It is used as the ending distribution $\rho_0$.}\label{fig2}
\end{figure}

In the experiments of this paper, the starting point is set to be the monthly averaged SSH in August, 1994 and the end point is set to be the monthly averaged SSH in August, 2014. The goal for the experiment is to simulate the optimal evolution path from the starting point to the end. In order to monitor the simulation results in detail, results in the northern Pacific region will be viewed and analyzed. This region is specially selected because the existence of Kuroshio current. Similar to the Gulf Stream in the North Atlantic, the Kuroshio Extension is a dynamic but relatively unstable system, with variability in the associated bifurcation latitude occurring on interannual time scales. The cause of these variations and their effects on the surface flow and total transport of waters has been studied extensively, SSH can be a direct indicator for Kuroshio so with recent advances in sea surface height satellite altimetry methods allowing for observational studies on larger timescales. As can be seen in figure \ref{fig1} and figure \ref{fig2}, both SSH patterns clearly indicate the Kuroshio current and its extension. 


\subsection{Results}
\subsubsection{Evolution of SSH}
A total of 350 iterations ($nk=350$) have been performed and inside of each iteration 51 steps were carried out to simulate the optimal path from the starting point (monthly averaged SSH in August, 1994) to the end point (monthly averaged SSH in August, 2014). Simulation results are saved for further analysis after each step inside of each iteration. 

\begin{figure}
\includegraphics[scale=0.23]{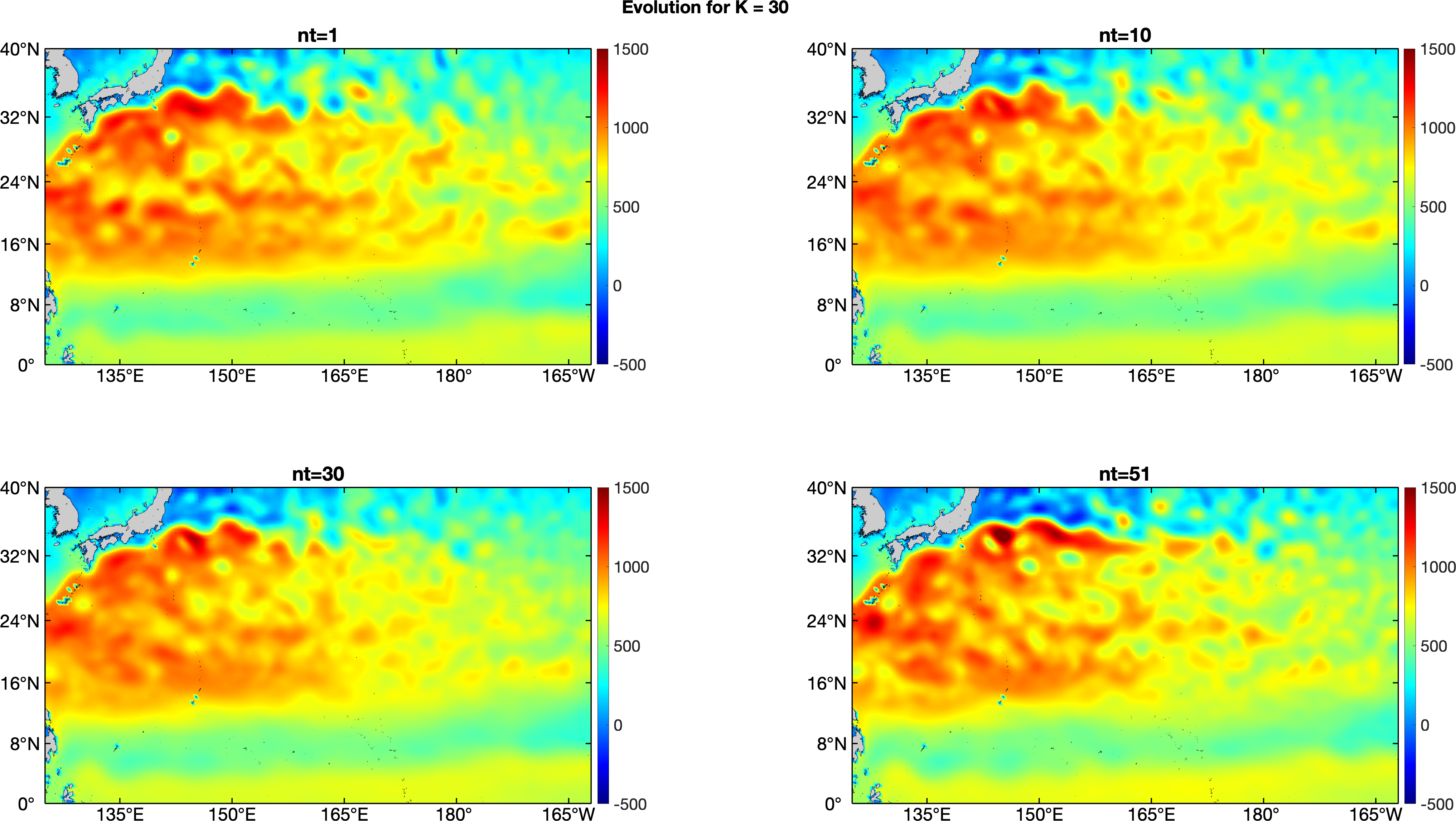}
\caption{Evolution of SSH at different time steps $nt=1,10,30,51$ after the iteration $nk=30$.} \label{fig3}
\end{figure}

\begin{figure}
\includegraphics[scale=0.23]{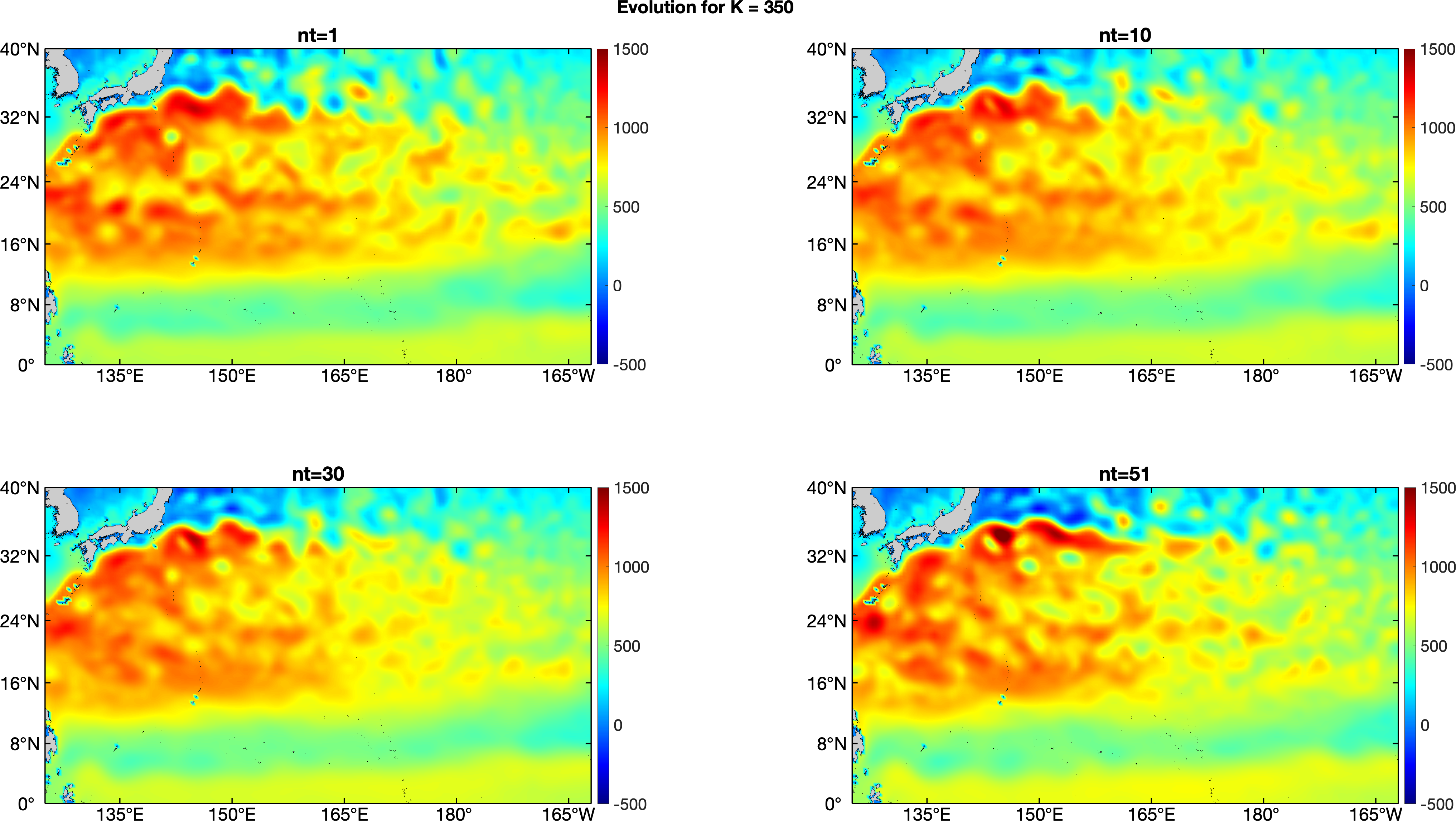}
\caption{Evolution of SSH at different time steps $nt=1,10,30,51$ after the iteration $nk=350$.} \label{fig4}
\end{figure}

In order to evaluate the performance of this algorithm, several comparisons are put forward. First, the SSH evolutions after different iterations are compared. Figure \ref{fig3} shows the SSH evolution after the 1st, 10th, 30th, and 51st steps after the 30th iteration. The process clearly illustrates the SSH evolution in the region of North Pacific, especially around the Kuroshio current area. The overall trend of SSH evolution revealed in the experiment is consistent with reality. As introduced in the introduction, the global warming trend in recent decades notably causing the continuous rising of the sea level. The enhancement of SSH mainly occurs on the right side of Kuroshio because of the Coriolis force. Note that several high SSH concentration areas which are caused by the convergence of ocean mesoscale eddies as well as low SSH concentration areas which are caused by the divergence of ocean mesoscale eddies are clearly indicated during the evolution process. The movements of those eddies can be clearly described in the evolution process as well. Figure \ref{fig4} shows the results of SSH evolution after the final iteration when $nk=350$. Similar as shown in figure \ref{fig3}, it also clearly indicates the evolution process for SSH, especially around Kuroshio area as well as the convergence and divergence caused by ocean mesoscale eddies. Note that although the difference between figure \ref{fig3} and figure \ref{fig4} is insignificant, it will indicate how this algorithm performs in terms of approaching the optimal evolution path. We will have more discussions in next session. 

\subsubsection{Comparison between different iteration steps}

From discussions in last section, we can conclude the capacity of this algorithm to simulate the evolution process for SSH. In this section, we will discuss the capacity of the algorithm to simulate the optimal path from starting point to the end point. As mentioned in Section 2, the optimal velocity field will be obtained as the iteration put forward. In another word, the optimal path for the SSH field from starting point (August, 1994) to the end (August, 2014) will be established as iteration goes on. Figure \ref{fig5} presents a sketch on how the optimal path is gained with iteration and approaching the “Actual Path” from the starting initial state to the end state. When the algorithm is working fine, the evolution path from initial to end will be optimized towards the direction of “Actual Path”. That means the difference between evolution path will be smaller as the iteration goes on.

\begin{figure}
 \includegraphics[scale=0.23]{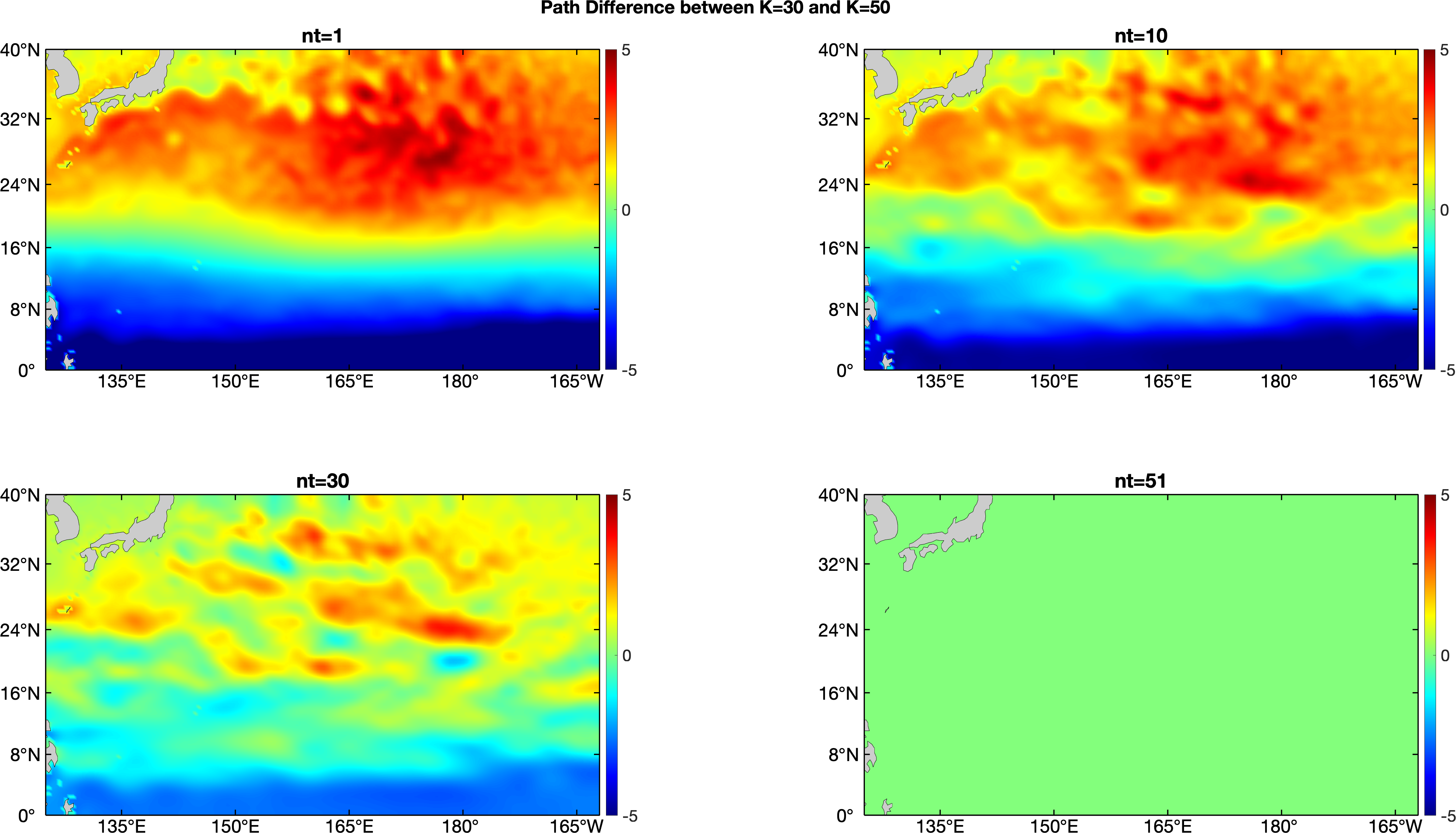} 
 \caption{The density difference of SSH evolution at time step $nt=1,10,30,51$ between the iteration $nk=30$ and $nk=50$.}\label{fig6}
 \end{figure}

\begin{figure}
 \includegraphics[scale=0.23]{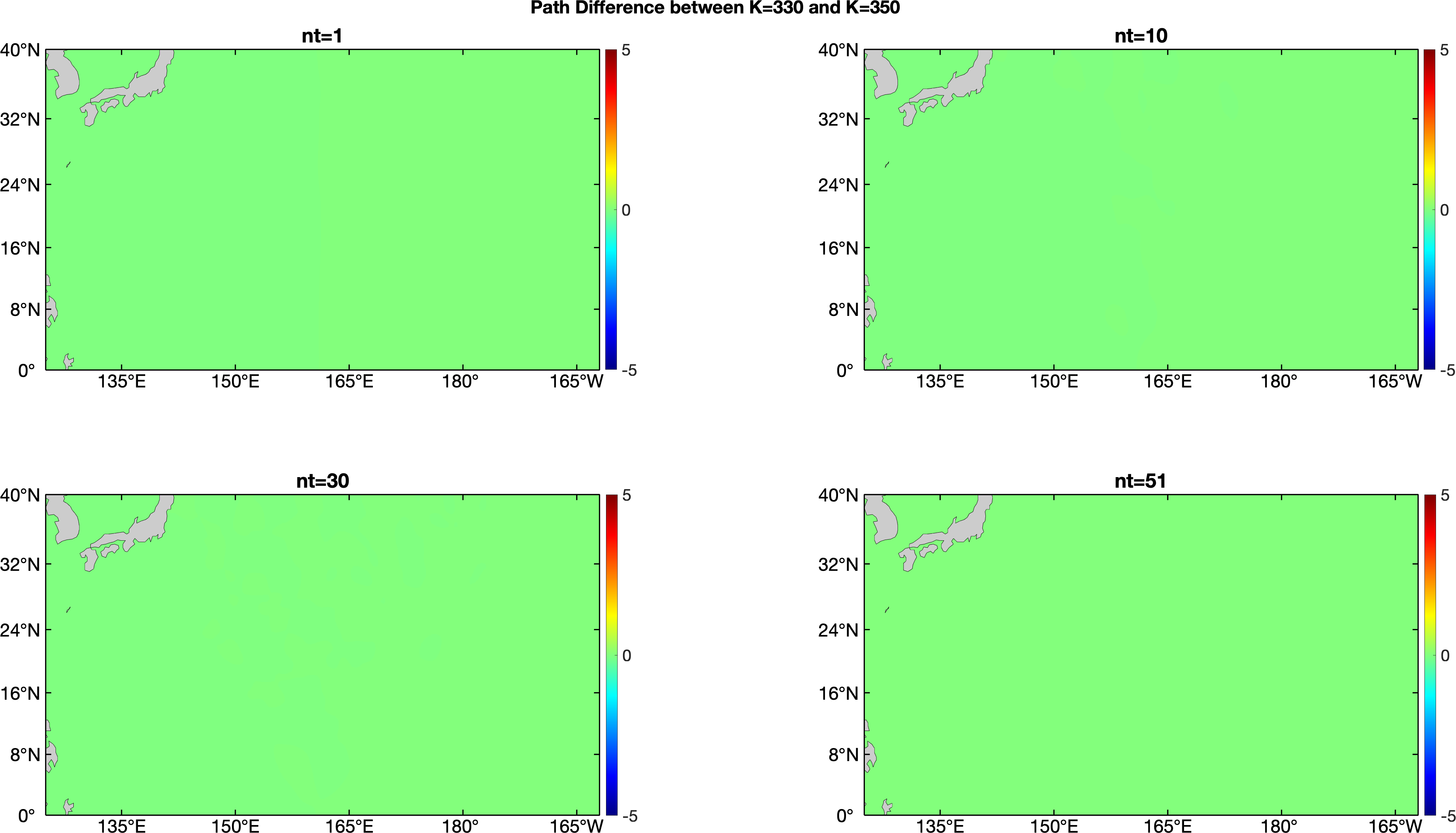} 
 \caption{The density difference of SSH evolution at time step $nt=1,10,30,51$ between the iteration $nk=330$ and $nk=350$.}\label{fig7}
 \end{figure}

The difference of SSH evolution process between iterations is calculated and analyzed as are shown in figure \ref{fig6} and figure \ref{fig7}. Figure \ref{fig6} shows the results between iteration 30 and 50. It can be noticed that although the difference between the starting status (not shown) and the end status are invisible from the patterns, the in-between process shows significant differences, indicating the different evolution path as shown in figure \ref{fig5}. However, at the end of the iterations, we analyzed the above differences between iteration 330 and 350 as shown in figure \ref{fig7}. It can be clearly noticed that the differences all fall below a small value with most part has a difference smaller than the machine precision. This is straightforward evidence indicating the evolution paths between the iterations being identical. According to the theory of optimal transport, the optimal velocity field has been reached. 

 
 \section{Discussion}
In this paper, the evolution process of SSH in the North Pacific is simulated based on a generalized Schr\"odinger bridge and Fokker-Planck solver. The evolution of SSH from August 1994 to August 2014 has been revealed. Evolution process indicates how sea level rising under the circumstances of global warming. The SSH evolutions in several significant areas induced by ocean mesoscale eddies are simulated. The underlying dynamic mechanisms for those evolution should be analyzed but it's beyond the scope of this paper. The optimal path and the convergence of the proposed iterative mixed control algorithm is then numerically verified by comparing the differences between results of different iterations. Clear evidences show the approaching of optimal path from starting status to the end status, indicating the success of the algorithm and its capability to find the optimal velocity field in optimal transport. We remark this type of finding the optimal transformation path between two given data is different from directly solving a gradient flow system \cite{gao2019gradient, gao2019global} in a Hilbert space or a Banach space because the transformation is finished in a finite time horizon.

\bibliographystyle{plain}
\bibliography{warm_bib}

\begin{thebibliography}{10}

\bibitem{CW2011}
John~A Church and Neil~J White.
\newblock Sea-level rise from the late 19th to the early 21st century.
\newblock {\em Surveys in geophysics}, 32(4):585--602, 2011.

\bibitem{cuturi2013sinkhorn}
Marco Cuturi.
\newblock Sinkhorn distances: Lightspeed computation of optimal transport.
\newblock {\em Advances in neural information processing systems}, 26, 2013.

\bibitem{Ishii_Evans_1985}
L.C. Evans and H.~Ishii.
\newblock A pde approach to some asymptotic problems concerning random
  differential equations with small noise intensities.
\newblock {\em Annales de l’Institut Henri Poincaré C, Analyse non
  linéaire}, 2(1):1–20, Feb 1985.

\bibitem{eymard2000finite}
Robert Eymard, Thierry Gallou{\"e}t, and Rapha{\`e}le Herbin.
\newblock Finite volume methods.
\newblock {\em Handbook of numerical analysis}, 7:713--1018, 2000.

\bibitem{FW}
Mark~I. Freidlin and Alexander~D. Wentzell.
\newblock {\em Random Perturbations of Dynamical Systems}, volume 260 of {\em
  Grundlehren der mathematischen Wissenschaften}.
\newblock Springer Berlin Heidelberg, 2012.

\bibitem{gao2021large}
Yu~Gao, Yuan Gao, and Jian-Guo Liu.
\newblock Large time behavior, bi-hamiltonian structure, and kinetic
  formulation for a complex burgers equation.
\newblock {\em Quarterly of Applied Mathematics}, 79(1):55--102, 2021.

\bibitem{gao2019global}
Yuan Gao.
\newblock Global strong solution with bv derivatives to singular solid-on-solid
  model with exponential nonlinearity.
\newblock {\em Journal of Differential Equations}, 267(7):4429--4447, 2019.

\bibitem{gao2021inbetweening}
Yuan Gao, Guangzhen Jin, and Jian-Guo Liu.
\newblock Inbetweening auto-animation via fokker-planck dynamics and
  thresholding.
\newblock {\em Inverse Problems \& Imaging}, 15(5):843, 2021.

\bibitem{gao2022transition}
Yuan Gao, Tiejun Li, Xiaoguang Li, and Jian-Guo Liu.
\newblock Transition path theory for langevin dynamics on manifold: optimal
  control and data-driven solver.
\newblock {\em Multiscale Modeling and Simulation}, 2022.

\bibitem{gao2018new}
Yuan Gao, Jin Liang, and Ti-Jun Xiao.
\newblock A new method to obtain uniform decay rates for multidimensional wave
  equations with nonlinear acoustic boundary conditions.
\newblock {\em SIAM Journal on Control and Optimization}, 56(2):1303--1320,
  2018.

\bibitem{yg20}
Yuan Gao and Jian-Guo Liu.
\newblock A note on parametric bayesian inference via gradient flows.
\newblock {\em Annals of Mathematical Sciences and Applications},
  5(2):261--282, 2020.

\bibitem{gao2022revisit}
Yuan Gao and Jian-Guo Liu.
\newblock Revisit of macroscopic dynamics for some non-equilibrium chemical
  reactions from a hamiltonian viewpoint.
\newblock {\em Journal of Statistical Physics}, 189(2):1--57, 2022.

\bibitem{gao2019gradient}
Yuan Gao, Jian-Guo Liu, and Xin~Yang Lu.
\newblock Gradient flow approach to an exponential thin film equation: global
  existence and latent singularity.
\newblock {\em ESAIM: Control, Optimisation and Calculus of Variations}, 25:49,
  2019.

\bibitem{gao2023data}
Yuan Gao, Jian-Guo Liu, and Nan Wu.
\newblock Data-driven efficient solvers for langevin dynamics on manifold in
  high dimensions.
\newblock {\em Applied and Computational Harmonic Analysis}, 62:261--309, 2023.

\bibitem{peyre2019computational}
Gabriel Peyr{\'e}, Marco Cuturi, et~al.
\newblock Computational optimal transport: With applications to data science.
\newblock {\em Foundations and Trends{\textregistered} in Machine Learning},
  11(5-6):355--607, 2019.

\bibitem{scharfetter1969large}
Donald~L Scharfetter and Hermann~K Gummel.
\newblock Large-signal analysis of a silicon read diode oscillator.
\newblock {\em IEEE Transactions on electron devices}, 16(1):64--77, 1969.

\bibitem{Schrodinger1932theorie}
Erwin Schr{\"o}dinger.
\newblock Sur la th{\'e}orie relativiste de l'{\'e}lectron et
  l'interpr{\'e}tation de la m{\'e}canique quantique.
\newblock In {\em Annales de l'institut Henri Poincar{\'e}}, volume~2, pages
  269--310, 1932.

\bibitem{sinkhorn1964relationship}
Richard Sinkhorn.
\newblock A relationship between arbitrary positive matrices and doubly
  stochastic matrices.
\newblock {\em The annals of mathematical statistics}, 35(2):876--879, 1964.

\bibitem{yasue1981stochastic}
Kunio Yasue.
\newblock Stochastic calculus of variations.
\newblock {\em Journal of functional Analysis}, 41(3):327--340, 1981.

\end{thebibliography}

\end{document}